\documentclass[12pt,preprint]{aastex}
\usepackage{amsmath}

\shorttitle{WD \& SNeIa}
\shortauthors{Ablimit at el.}

\begin{document}


\title{Evolution of Magnetized White Dwarf Binaries to Type Ia Supernovae}
\author{Iminhaji Ablimit\altaffilmark{1,\star} and
Keiichi Maeda\altaffilmark{1} }
\altaffiltext{}{$^\star$Corresponding author;  JSPS International Research Fellow; iminhaji@kusastro.kyoto-u.ac.jp}
\altaffiltext{1}{Department of Astronomy, Kyoto University, Kitashirakawa-Oiwake-cho, Sakyo-ku, Kyoto 606-8502, Japan}


\begin{abstract}

With the increasing number of observed magnetic white dwarfs (WDs), the role of magnetic field of the WD in both single and binary evolutions should draw more attentions. In this study, we investigate the WD/main-sequence star binary evolution with the Modules for Experiments in Stellar Astrophysics (MESA code), by considering WDs with non-, intermediate and high magnetic field strength. We mainly focus on how the strong magnetic field of the WD (in a polar-like system) affects the binary evolution towards type Ia supernovae (SNe Ia). The accreted matter goes along the magnetic field lines and falls down onto polar caps, and it can be confined by the strong magnetic field of the WD, so that the enhanced isotropic pole-mass transfer rate can let the WD grow in mass even with a low mass donor with the low Roche-lobe overflow mass transfer rate. The results under the magnetic confinement model show that both initial parameter space for SNe Ia and characteristics of the donors after SNe Ia are quite distinguishable from those found in pervious SNe Ia progenitor models. The predicted natures of the donors are compatible with the non-detection of a companion in several SN remnants and nearby SNe.

\end{abstract}


\section{Introduction}

The evolution of a white dwarf (WD) binary to a type Ia supernova (SN Ia) has been widely studied. There are however still open questions about the WD binary evolution and origin of SNe Ia. There are two popular scenarios leading to the thermonuclear explosion of a WD as an SN Ia (Hoyle \& Fowler 1960; Webbink 1979; Nomoto 1982; Iben \& Tutukov 1984; Webbink 1984; Hachisu et al. 1996; Ablimit, Maeda \& Li  2016). One is the single degenerate  (SD) model, in which a CO WD accumulates mass from a non-degenerate donor star. In the other model, the so-called double degenerate (DD), the merger of two WDs in a binary leads to an SN Ia. Some other scenarios have also been proposed including the core-degenerate scenario in which a merger of a WD with the core of an asymptotic giant branch star or a red giant star may cause an SN Ia during the final stage of the common envelope evolution (Soker 2011;  Kashi \& Soker 2011). However,  it is still under debate as for which scenario contributes to the observed SNe Ia (e.g., Howell 2011;  Maeda \& Terada 2016; Livio \& Mazzali 2018).

In the SD scenario, the mass transfer and accretion (mass retention efficiency) are key issues (Kahabka \& van den Heuvel 1997). To realize the steady burning and let a WD growth in mass (Nomoto et al. 2007; Shen \& Bildsten 2007), the material should be transferred from a relatively massive main-sequence (MS) star or a low mass (sub)giant star on the thermal or nuclear timescale (Rappaport et al. 1994; Hachisu et al. 1996; Li \& van den Heuvel 1997; Yungelson et al. 1996; Wang et al. 2010). As a consequence, the SD generally predicts the existence of a non-degenerate companion star at or after the SN explosion of the primary WD. However, no such a companion star has been directly identified yet. For some nearby SNe and SN remnants (SNRs), upper-limits on the luminosity of the companion star have been used to reject some parameter space in the SD scenario (Maoz, Mannucci \& Nelemans 2014; Ruiz-Lapuente 2014).

However, it is still too early to conclude that the SD scenario cannot be a major pathway toward SNe Ia. It is possible that there are some important physical processes so far missing in modeling the evolution of mass-accreting WD binaries. Indeed, different groups have differently treated the mass transfer and accretion processes to investigate the outcome of the SD scenario, and the outcome is sensitive to this treatment  (e.g., Hachisu et al. 2008; King \& van Teeseling 1998; Ablimit, Xu \& Li 2014). Also, observations of some supersoft X-ray sources (SSSs) are in tension with the current WD binary evolution models (Ablimit, Xu \& Li 2014; Ablimit \& Li 2015). These suggest that there could be a missing process in the current models of the accreting WD binary evolution, and this might affect the parameter space of the SD scenario  toward SNe Ia.

Around $10\%$ of the observed WDs are estimated to have the magnetic field strengths from $10^3$ to a few$\times10^8$ Guess in volume-complete samples (Liebert et al. 2003; Schmidt et al. 2003; Kawka et al. 2007). The role of magnetic field in the WD binary evolution has not been sufficiently explored, while it has been pointed out that the magnetic field has crucial effects on the evolution of WD binaries (e.g., Ablimit, Xu \& Li 2014; Farihi et al. 2017). Cataclysmic variables (CVs) are another system of interacting WD binaries undergoing the mass transfer from a MS star or brown dwarf star to the WD. Around $25\%$ of all known CVs are magnetic CVs (Ferrario et al. 2015). They are divided into non-magnetic CVs ($<$ 1 MG), intermediate polars ($\sim$1--10 MG) and polars ($> $10 MG) depending on the magnetic field strength of the WD, and the observed WD magnetic field strength so far is up to 230 MG (Schmidt et al. 1999).  Magnetic WDs in SSS and symbiotic binaries have been discovered as well (Kahabka 1995; Sokoloski \& Bildsten 1999; Osborne et al. 2001).  The mean mass of highly magnetized WDs is $\sim 0.8\rm M_\sun$ which is significantly higher than the mean mass of non-magnetic WDs, $\sim 0.6\rm M_\sun$ (Kepler et al. 2013). This implies that the evolution and mass growth of magnetic WDs are different from those of non-magnetic WDs.

 Norton et al. (2008) discussed the spin-orbit equilibrium mediated by the magnetic field for the intermediate polar system, and applied the results to the CV evolution. One possible effect of the magnetic field toward SNe Ia has been recently discussed by Neunterfel et al. (2017), who examined the helium accretion onto a weakly magnetized CO WD, taking into account the angular momentum transport by the magnetic field. They claimed that this kind of evolution could lead to fast and faint SNe Ia rather than classical SNe Ia. The effect of the strong magnetic field on the WD binary accretion process, however, is largely unexplored. For the magnetic WD binaries, Livio (1983) proposed that the accreted matter can be confined in the polar regions if the WD in the binary has sufficiently strong magnetic field strength (see also Wheeler 2012).  Cumming (2002) found the ohmic diffusion timescale of the magnetic field of the WD is much longer than the accretion timescale. Therefore, the accreted mass is expected to be confined in the polar caps.

In this paper, we first specifically address the magnetic confinement model. We investigate possible influence of magnetic field of the WD during the binary evolution, and present detailed numerical calculations of the mass transfer process. In \S 2, we describe our models to treat the binary evolution processes including the magnetic confinement. The results are presented in \S 3. The paper is closed in \S 4 with conclusions and discussion.

\section{The WD binary evolution with the magnetic confinement}
\label{sec:model}

We calculate the WD binary evolution including a MS donor star with the initial masses ($M_{\rm d}$) ranging from 0.8 to 8 $\rm M_\sun$ by using the version (10000) of Modules for Experiments in Stellar Astrophysics (MESA; Paxton et al. 2011, 2013, 2015). The initial orbital period ranges from 0.1 to 30 days. Within the parameter space, the donor mass and orbital period are discretized with bins of 0.1 $\rm M_\sun$ and 0.2 days, respectively. The WD is approximated as a point mass. A typical Population I composition with H abundance X = 0.70, He abundance Y = 0.28 and metallicity Z = 0.02 is taken for the donor star.  The formula used for the effective Roche-lobe (RL) radius of the donor star (Eggleton1983) is,

\begin{equation}
\frac{R_{\rm L, d}}{a} = \frac{0.49q^{2/3}}{0.6q^{2/3} + {\rm ln}(1+q^{1/3})} ,
\end{equation}
where $q={M_{\rm d}}/{M_{\rm WD}}$ and $a$ is the orbital separation. The Ritter scheme (Ritter 1988; Paxton et al. 2015) is adopted to calculate the mass transfer via RL overflow in the code. The angular momentum loss caused by gravitational wave radiation (Landau \& Lifshitz 1975)  and magnetic braking of the companion (Verbunt \& Zwaan 1981; Rappaport et al. 1983) is also included. Other stellar and binary evolution parameters are fixed to be as the typical ones introduced in the MESA instrumental papers (e.g., Paxton et al. 2015).

In order to see the effect of the high magnetic field of the WD, we consider non-magnetic and magnetic WDs with the initial masses ($M_{\rm WD,i}$) of $0.8\rm M_\sun$, $1.0\rm M_\sun$ and $1.2\rm M_\sun$. In the magnetic WD binaries, the magnetic field strengths of WDs are assumed as B $=1.53\times10^8$, $1.10\times10^8$ and  $2.00\times10^6$ Gauss. We assume the WD explodes as an SN Ia when it grows in mass and reaches to the Chandrasekhar limit mass ($\rm M_{\rm Ch} =$1.38$\rm M_\sun$). Accreting WDs may rotate differentially, and they may not reach the central carbon ignition condition even when they grow beyond the canonical Chandrasekhar limit (Yoon \& Langer 2004). However, we assume all WDs explode when their mass reach to the Chandrasekhar limit in order to highlight the role of the magnetic field in the SD scenario.

\subsection{The magnetic confinement during the mass transfer}

If the magnetic field is not strong enough, an accretion disk around the WD can be formed after the donor fills its RL and starts the mass loss through Lagrangian point 1. In the polar-like systems, the strong magnetic field of the WD lets the binary have a synchronous rotation and prevent the accretion disk formation (Cropper 1990). The stream of matter from the donor is captured by the magnetic field of the WD, then it follows the magnetic field lines and falls down onto the magnetic poles of the WD as an accretion column (see Figure 1).  Livio (1983) showed that the magnetic confinement can suppress the nova burst even at the low mass transfer rate. Livio (1983) also pointed out that the critical value of magnetic field strength to realize the magnetic confinement depends on the mass of the WD.

The appropriate physical condition for the magnetic confinement is that the magnetic field strength B of the WD satisfies the following condition (Spitzer 1962; Livio 1983),

\begin{equation}
\rm{B} \geq 9.3\times10^{7} (\frac{\rm{R_{WD}}}{5\times10^8\,\rm cm})(\frac{\rm{P_b}}{5\times10^{19}\,\rm{dyne\,cm^{-2}}})^{7/10}(\frac{\rm{M_{WD}}}{\rm M_\sun})^{-1/2}(\frac{\dot{M}}{10^{-10}\,\rm{M_\sun\,yr^{-1}}})^{-1/2}\,,
\end{equation}
where $\dot{M}$ is the RLOF mass transfer. $\rm{P_b}$ is the pressure at the base of accreted matter.
For the mass ($\rm{M_{WD}}$) -- radius ($\rm{R_{WD}}$) relation of the WD, the following equation is used (Nauenberg 1972),

\begin{equation}
\rm{R_{WD}} = 7.8\times10^8[(\frac{\rm{M_{WD}}}{\rm{M_{Ch}}})^{-2/3} - (\frac{\rm{M_{WD}}}{\rm{M_{Ch}}})^{2/3}]^{1/2} ,
\end{equation}
where $\rm{M_{Ch}}$ is the Chandrasekhar limit mass.
Accreted matter will attempt to diffuse perpendicularly to the magnetic field lines and spread over the surface of the WD, and the condition (2) shown above means that the diffusion timescale of the accreted matter is longer than the time required to build up the necessary pressure for an outburst (see Woosely \& Wallace 1982). The stable hydrogen burning would occur instead of the outburst if the magnetic field strength of the WD fulfills the condition (2).

The strong magnetic field confinement can inhibit the nova outburst, according to the following arguments by Livio (1983). The nova outburst is expected to occur if the mass of accreted material is sufficiently large when $\rm{P_b}$ reaches to a certain critical value (a few$\times10^{19}\,\rm{dyne\,cm^{-2}}$) to initiate the hydrogen burning (Fujimoto 1982a, b; Livio 1983).
This condition is transferred to a critical mass transfer rate, below which the nova outburst takes place (Kato, Hachisu \& Saio 2017). If the accreted matter is confined to the polar column, then the following isotropic pole-mass transfer rate ($\dot{M}_{\rm{p}}$) should be compared to the critical mass transfer rate, rather than the usual RLOF mass transfer rate ($\dot{M}$). The isotropic pole-mass transfer rate ($\dot{M}_{\rm{p}}$) is given as follows,

\begin{equation}
\dot{M}_{\rm{p}} =  \frac{\rm S}{\Delta \rm{S}} {\dot{M}},
\end{equation}
where $\rm S$ is the surface area of the WD, and $\Delta \rm{S}$ is that of the dipole regions. $\Delta \rm{S}$ can be calculated by the equations of dipole geometry and Alfv$\acute{\rm e}$n radius of the magnetic WD.  Livio (1983) showed that the isotropic pole-mass transfer rate should become always larger than the critical accretion rate, in case the magnetic field is so strong to satisfy the condition (2) for the pressure ($\rm{P_b}$) otherwise appropriate for the initiation of a nova outburst (i.e., $\sim 5\times10^{19}\,\rm{dyne\,cm^{-2}}$). The highly magnetized WD can therefore grow in mass without nova eruption (see \S 2.2).

\subsection{The mass growth of the magnetized WD}

The mass transfer rate is important to realize the stable hydrogen and helium burning on the WD, and it will affect the mass retention and growth of the WD. Kato, Hachisu \& Saio (2017) summarized possible effects which affect the mass retention efficiency and discussed results by several groups on this efficiency.  We adopt the prescription of the Prialniks group for the efficiency of hydrogen burning (see Prialnik \& Kovetz 1995; Yaron et al. 2005; Hillman et al. 2015, 2016), and adopt methods of Kato \& Hachisu (2004) for the mass accumulation efficiency of helium. The mass growth rate of the WD is,

\begin{equation}
\dot{M}_{\rm{WD}} =  \eta_{\rm H} \eta_{\rm He} {\dot{M}},
\end{equation}
where $\eta_{\rm H}$ and $\eta_{\rm He}$ are efficiencies of hydrogen and helium burning\footnote{The C/O ashes from the steady helium burning shell may unstably ignite in a shell flash, but these carbon burning flashes do not significantly affect the growth of the WD (Brooks et al. 2017).}, and $\dot{M}$ is the RLOF mass transfer rate to the WD. For the WD binary with non or weak magnetic field, the burning efficiencies are determined by the RLOF mass transfer rate ($\dot{M}$).
 For the WD binaries with the strong magnetic field, $\dot{M}$ in the prescriptions for  $\eta_{\rm H}$ and $\eta_{\rm He}$ should be replaced by the isotropic polar-mass transfer rate ($\dot{M}_{\rm{p}}$). The nova eruption will take place when the RLOF mass transfer rate is lower than $ \sim 10^{-7}\, M_\sun\,{\rm yr^{-1}}$ in the non-magnetic case, and this prevents the WD mass growth\footnote{In this paper, we investigate the condition of the steady-state burning. However, we note that there would be a net accretion in the nova systems as well (Hillman et al. 2015, 2016; Henze et al. 2018)}. In the case of highly magnetized WD binaries, the isotropic polar-mass transfer rate could be higher than $ \sim 10^{-7}\, M_\sun\,{\rm yr^{-1}}$ even with the low RLOF mass transfer rate. For example, $\eta_{\rm H}$ is (practically) 0 when the mass transfer rate $ \dot{M} \sim 10^{-9}\, M_\sun\,{\rm yr^{-1}}$  in the non or weak magnetic field case. However, under the high magnetic field case, $\dot{M}_{\rm p} > 10^{-7}\,\rm M_\sun\,{\rm yr^{-1}}$ even with the RLOF mass transfer rate $ \dot{M} \sim 10^{-9}\,\rm M_\sun\,{\rm yr^{-1}}$, and $\eta_{\rm H} \sim 1 $. Therefore, the magnetized WD can grow in mass and reach to $\rm M_{\rm Ch}$ in the magnetic confinement model.

\section{Results }

\subsection{The effect of the magnetic field in the WD binary evolution}

We select some examples to show the binary evolution with and without the magnetic confinement under different initial conditions. Figure 2 shows a WD binary evolution ($\rm M_{\rm WD, i} = 1.2\rm M_\sun$, $\rm M_{\rm donor, i} = 1.2\rm M_\sun$ and initial orbital period is 10 days) without and with the magnetic field. In the non-magnetic case, the RLOF mass transfer rate in this binary is too low to let the WD grow in mass, and it evolves as the typical CV with the period gap (see Paxton et al. 2015). In the early phase before $\rm{log}_{10} t \sim 9.78212$, the angular momentum loss due to the magnetic braking and RLOF mass transfer rate in both cases are basically same. However, in the later evolution after $\rm{log}_{10} t \sim 9.78212$, the non-magnetic WD binary continues the RLOF mass transfer while the magnetic WD binary terminates the evolution due to the SN explosion under the magnetic confinement. In the magnetic case, the magnetic field plays crucial role, the enhanced polar mass transfer rate realizes the stable burning even with the low RLOF mass transfer and nova burst can be avoided, and the WD efficiently accretes the mass from the donor. Thus, the magnetic WD reaches $\rm M_{\rm Ch}$ and explodes as SNe Ia, and there is no angular momentum loss due to the mass loss during the evolution (Figure 2). The orbital period evolves from 10 days to 100 days in the non-magnetic case, because the system loses angular momentum further in the later evolution due to the mass loss (nova burst happens and no significant accretion) (Figure 2). Because of the mass loss, the donor mass in the non-magnetic case evolves further toward a lower mass than the final donor mass at an SN Ia in the magnetic case. Thus, there is some difference in other properties (surface gravity and radius) of the donors in the late phases of two cases (see Figure 2).

The evolution toward SNe Ia under the magnetic confinement scenario depends on the WD mass (Figures 2 \& 3). The evolutions of $\rm M_{\rm WD, i} = 0.8\rm M_\sun$ with the initial orbital period of 1 day and  $\rm M_{\rm WD, i} = 1.0\rm M_\sun$ with the initial orbital period of 1.2 days under the magnetic confinement model are shown in Figure 3. The initial donor stars' masses are 1.2$\rm M_\sun$ and 2.2$\rm M_\sun$, respectively. If these less massive WDs in Figure 3 have large initial orbital periods, e.g., $P_{\rm orb,i} = 10$ days for $\rm M_{\rm WD, i} = 1.2\rm M_\sun$ as shown in Figure 2, there would be practically no mass exchange through the RLOF. In the first binary (left panels in Figure 3), the traditional thermal timescale mass transfer is not high enough for the stable H burning to take place. However, the polar mass rate is enhanced up to a few $10^{-7}\,\rm M_\sun\,{\rm yr^{-1}}$ by the magnetic confinement, thus the WD can grow in mass to $\rm M_{\rm Ch}$.

In the binary with $\rm M_{\rm WD, i} = 1.0\rm M_\sun$ in Figure 3, the donor mass is much larger than the WD mass. Although the magnetic confinement also works, the mass transfer actually proceeds on a thermal timescale, similar as in typical supersoft X-ray binaries (e.g., Li \& van den Heuvel 1997). With the higher mass donor  (higher RLOF mass transfer rate), the polar mass transfer rate will be higher, and the outcomes will be different accordingly. Thus,  the donor mass also plays important role in the evolution. The orbital evolution is dominated by the mass loss during the earlier time of the RLOF mass transfer (see right panels of Figure 3). The gravitational wave radiation dominates the orbital evolution in the late time of the evolution in both binaries. The binary with $\rm M_{\rm WD, i} = 0.8\rm M_\sun$ could not have the confinement with the same magnetic field strength used for the binary with $\rm M_{\rm WD, i} = 1.0\rm M_\sun$.  The stronger magnetic field is required to realize the pole mass transfer for the less massive WD (the condition (2) in \S 2.1).

\subsection{Implications for SNe Ia}

The initial donor mass and orbital period distributions to produce SNe Ia are given in Figure 4. In the figure, the results of $\rm M_{\rm WD, i} = 0.8\rm M_\sun$ with B $=1.53\times10^8$ G,  $\rm M_{\rm WD, i} = 1.0\rm M_\sun$ with B $=1.1\times10^8$ G,  $\rm M_{\rm WD, i} = 1.2 \rm M_\sun$ with B $=1.1\times10^8$ G, and the case of non-magnetic WD with  $\rm M_{\rm WD, i} = 1.2 \rm M_\sun$ are showed. The range of initial donor mass and orbital period are 0.8--2.2$\rm M_\sun$ with 0.3--3.2 days, 0.8--2.7$\rm M_\sun$ with 0.3--4.0 days and
0.8--3.2$\rm M_\sun$ with 0.3--25 days, for those magnetic WDs with $\rm M_{\rm WD, i} = $0.8, 1.0 and $1.2 \rm M_\sun$, respectively. For the non-magnetic WD with $\rm M_{\rm WD, i} = 1.2 \rm M_\sun$, it is 2.2--3.4$\rm M_\sun$ and 0.4--4.5 days, being consistent with typical values of previous works (e.g., Li \& van den Heuvel 1997). As compared to the non-magnetic case and previous works, the ranges of initial orbital period and initial donor mass are larger for the high magnetic field case, because the magnetic confinement leads to the enhanced isotropic pole-mass transfer, and the accreted matter can burn stably with this higher pole-mass transfer even with the low mass donor star. The upper limits of the regions in Figure 4 are based on the occurrence of common envelope if the isotropic pole-mass transfer rate is higher than $10^{-4}\,\rm M_\sun\,\rm{yr^{-1}}$, and the lower limits are determined by the initiation of novae burst if the polar-mass transfer is lower than $\sim 10^{-7}\,\rm M_\sun\,\rm{yr^{-1}}$. The left and right limits in the Figure 4 are derived according to the mass-radius relation of zero age MS stars and full exhaustion of the central hydrogen, respectively.
It is worth noting that the initial orbital period can be up to 25 days when the initial donor mass is lower than 1.8$\rm M_\sun$ (Figure 2). If a less WD mass is coupled with long orbital period, the system keeps detached during the evolution, thus there is no mass exchange that the high magnetic field could affect.

The results from different values of magnetic field strength on $\rm M_{\rm WD, i} = 1.0 \rm M_\sun$  are shown in the right panel of Figure 4. The WD binaries with lower magnetic field strengths (intermediate polar-like systems) have the same initial parameter space as the non-magnetic WD binaries. This is because the lower magnetic field strength is not sufficiently strong to have the magnetic confinement transfer, therefore the intermediate polar-like system behaves in the same way with the non-magnetic WD.

In Figure 5, we show the distributions of the properties of the WD binaries and the donor stars at the time of the SN explosion. If we select a binary with $\rm M_{\rm WD, i} = 0.8 \rm M_\sun$, $\rm M_{\rm 2, i} = 1.0 \rm M_\sun$ and $\rm P_{\rm orb, i} = 0.3\,day$ from the initial space distributions in the Figure 4, the corresponding results in the Figure 5 are $\rm P_{\rm orb, f} \sim 0.132\,day$, $\rm M_{\rm 2, f} \sim 0.366 \rm M_\sun$,  $log_{10}$(radius/$R_\sun$)$\sim -0.448$, $log_{10}$($g$)$\sim 4.899$, $log_{10}$($\rm T_{\rm eff}$)$\sim 3.752$ and $\rm M_{\rm V}\sim 8.99$, respectively.  Combining the results of $\rm M_{\rm WD, i} = 0.8$ and $1.2 \rm M_\sun$, the ranges of final donor mass, $log_{10}$(radius/$R_\sun$), $log_{10}$(surface gravity ($g$)) and $log_{10}$(effective temperature ($\rm T_{\rm eff}$)) are $\sim 0.1$ -- 1.42 $\rm M_\sun$, -0.77--1.4, 1.43--5.2 and 3.5--3.83, respectively. The final orbital period distribution ranges from 0.07 to 61 days. The absolute magnitude of the donors ranges from -0.42 to 11 mag. The donor can be as dim as having 11 mag, since the less massive donor can now lead to an SN Ia. Such a faint donor is difficult to detect, and indeed is fainter than most (or all) of current observational upper limits on the donor's brightness for nearby SNe and SNRs (see below).

SN 1572 (Tycho's Supernova) still remains controversial about its surviving companion (e.g, Ruiz-Lapuente et al. 2004; Fuhrmann 2005; Ihara et al. 2007). From our results it might be possible that the surviving companion star could be too dim to detect. There are observational limits that the absolute magnitudes of any possible companion stars associated with SN 2011fe /PTF11kly and SN 1006 must be fainter than $\rm M_{\rm V} = 4.2$ and $4.9$ mags, respectively (Li et al. 2011; Edwards et al. 2012; Gonz$\acute{\rm a}$lez-Hern$\acute{\rm a}$ndez et al. 2012). Schaefer \& Pagnotta (2012) claimed by using the HST deep images that the central area of SNR 0509-67.5 in the LMC is empty of point sources down to $\rm{M_v}=8.4$ mag. The aforementioned observational constraints on the possible surviving companions have strong tension to what is expected by the traditional SD models, but those constraints can be consistent with the SD scenario once the effect of strong magnetic field is taken into account as shown in this work.

\section{Conclusions and Discussion}

With the low mass transfer rate ($<\,\sim10^{-7}$ $\rm M_\sun\,\rm{yr}^{-1}$), WDs barely grow to the Chandrasekhar mass limit because the most of the envelope is instantly ejected during nova outbursts. If ROLF mass transfer rate is too high ($>\,\sim10^{-6}$ $\rm M_\sun\,\rm{yr}^{-1}$), the mass transfer becomes dynamically unstable and the system evolves into a common envelope (e.g., Iben \& Livio 1993). To guarantee the proper mass transfer rate to realize the stable burning, the donor mass (mass ratio $> 5/6$) should be in the stable range. In the traditional SD model, the MS donors with $\sim 2 - 3.5 \rm M_\sun$ or the giant donors with $>$ 1.16 $\rm M_\sun$ (e.g., Li \& van den Heuvel 1997; Schaefer \& Pagnotta 2012) are required.

In this work, we calculate the WD and MS star binary evolution by assuming the WD has no, intermediate or high magnetic field. We mainly focus on the high magnetic field WD binary evolution by using MESA code under the magnetic confinement model (Livio 1983) to generate SNe Ia. In the case of non- and intermediate magnetic field of the WD, the binary evolution and the initial parameter space are the same as given in previous works. In the magnetic confinement model, the transferred mass falls down onto the polar caps along magnetic lines and can be confined by the high magnetic field. Then, the density of the accreted matter increases due to this confinement, and the nova criterion must be evaluated with this isotropic pole-mass transfer rate. This pole-mass transfer rate can be sufficiently large to let the accreted hydrogen burn stably even with low mass donor stars. We assume the timescale of the magnetic decay due to the mass accretion is at least an order of the thermal timescale. Thus, the high magnetic field of a WD could maintain its strength to confine the accreted matter (Livio 1983; Cumming 2002).

Under the magnetic confinement model, the initial parameter spaces for producing SNe Ia become larger than the previous studies without the effects of the magnetic field. Especially, the possible initial mass of donor stars extends to the lower mass region, as such systems can realize the stable burning in our scenario (\S 3.2). The final properties of donors derived in this paper are comparable with non-detection of (surviving) companion stars in nearby SNe and SNRs with the currently reported upper limits. As an extreme example, our model allows a donor as dim as 11 mag in the absolute $V$-band magnitude. We find that the delay time in our model is ranged from $\sim 10^8$ yr to $\sim 10^{11}$ yr. In the future, it is interesting to further investigate possible contribution of this kind of WD binaries to the whole population and diversity of SNe Ia.

\begin{acknowledgements}

This work was funded by JSPS International Postdoctoral fellowship of Japan (P17022), JSPS KAKENHI grant no.17F17022. The work was also supported by JSPS KAKENHI grant no. 17H02864, 18H04585 and 18H05223), Japan.
\end{acknowledgements}


REFERENCES

Ablimit, I., Xu, X.-J. \& Li, X.-D., 2014, ApJ, 780, 80

Ablimit, I.  \& Li, X.-D., 2015, ApJ, 815, 17

Ablimit, I., Maeda, K. \& Li, X.-D., 2016, ApJ, 826, 53

Brooks, J., Schwab, J., Bildsten, L., Quataert, E., \& Paxton, B., 2017, ApJ, 843,151

Cumming, A. 2002, MNRAS, 333, 589

Cropper M. 1990, Space Sci. Rew., 54, 195

Edwards, Z. I., Pagnotta, A. \& Schaefer, B. E. 2012, ApJL, 747, L19

Eggleton, P. P. 1983, ApJ, 268, 368

Farihi, J., Fossati, L., Wheatley, B. D. et al. 2017, arXiv:1709.08206v2

Ferrario, L., de Martino, D. \& Gsicke, B. T. 2015, SSRv, 191, 111F

Fuhrmann, K. 2005, MNRAS, 359, L35

Fujimoto, M. Y. 1982a, ApJ, 257, 752

Fujimoto, M. Y. 1982b, ApJ, 257, 767

Gonz$\acute{\rm a}$lez-Hern$\acute{\rm a}$ndez, J. I., Ruiz-Lapuente, P.,  Tabenero H. M., et al. 2012, Natur, 489, 533

Hachisu, I., Kato, M., \& Nomoto, K. 1996, ApJL, 470, L97

Hachisu, I., Kato, M., \& Nomoto, K. 2008, ApJ, 679, 1390

Hachisu, I., Kato, M., Saio, H. \& Nomoto, K. 2012, ApJ, 744, 69

Henze, M., Darnley, M. J., Williams, S. C., Kato, M., Hachisu, I. et al. 2018, ApJ, 857, 68

Hillman, Y., Prialnik, D., Kovetz A. \& Shara, M. M. 2015, MNRAS, 446, 1924

Hillman, Y., Prialnik, D., Kovetz, A. \& Shara, M. M. 2016, ApJ, 819, 168


Hoyle, F. \& Fowler, W. A., 1960, ApJ, 132, 565

Howell, D. A. 2011, NatCo, 2, 350

Iben, I. \& Tutukov, A. V., 1984, ApJS, 54, 335

Iben \& Livio 1993, PASP, 105, 1373

Ihara, Y., Ozaki, J., Doi, M., et al. 2007, PASJ, 59, 811

Kahabka, P. 1995, ASP Conference Series, Vol. 85

Kahabka, P., \& van den Heuvel, E. P. J. 1997, ARA\&A, 35, 69

Kato, M., \& Hachisu, I. 2004, ApJL, 613, L129

Kato, M., Hachisu, I. \& Saio, H. 2017, arXiv: 1711.01529v2

Kawaka, A., Vennes, S., Schmidt, G. D. et al., 2007, ApJ, 654, 499

Kashi, A. \& Soker, N. 2011, MNRAS, 417, 1466

Kepler, S. O., Pelisoli, I., Jordan, S., et al. 2013, MNRAS, 429, 2934

King, A. R. \& van Teeseling, A. 1998, A\&A, 338, 965

Landau, L. D., \& Lifshitz, E. M. 1975, The Classical Theory of Fields (4th ed.;
Oxford: Pergamon)

Li, W., Bloom, J. S., Podsiadlowski, Ph., et al. 2011, Natur, 480, 348

Li, X.-D., \& van den Heuvel, E. P. J. 1997, A\&A, 322, L9

Liebert, J., Bergeron, P., \& Holdberg, J. B. 2003, AJ, 125, 348

Livio, M. 1983, A\&A, 121, L7

Livio, M. \& Mazzali, P. 2018, PhR, 736, 1L

Maeda, K. \& Terada, Y. 2016, IJMPD, 253002

Maoz, D., Mannucci, F. \& Nelemans, G., 2014, ARA\&A, 52, 107


Nauenberg, M. 1972, ApJ, 175, 417

Neunteufel, P., Yoon, S.-C. \& Langer, N. 2017, A\&A, 602, 55

Nomoto, K., 1982, ApJ, 257, 780

Nomoto, K., Saio, H., Kato, M., \& Hachisu, I. 2007, ApJ, 663, 1269

Norton, A. J., Butters, O. W., Parker, T. L. \& Wynn, G. A. 2008, ApJ, 672, 524

Osborne, et al. 2001, A\&A, 378, 800

Paxton, B., Bildsten, L., Dotter, A., et al. 2011, ApJS, 192, 3

Paxton, B., Cantiello, M., Arras, P., et al. 2013, ApJS, 208, 4

Paxton, B., Marchant, P., Schwab, J., et al. 2015, ApJS, 220, 15

Prialnik, D. \& Kovetz A.1995, ApJ, 445, 789

Rappaport, S., Verbunt, F., \& Joss, P. C. 1983, ApJ, 275, 713

Rappaport, S., Di Stefano, F., \& Smith, J. D. 1994, ApJ, 426, 492

Ruiz-Lapuente, P., Comeron, F., Mndez, J., et al. 2004, Natur, 431, 1069

Ruiz-Lapuente, P., 2014, New Astron. Rev., 62, 15

Ritter, H. 1988, A\&A, 202, 93

Scalzo, R. A., Aldering, G., Antilogus, P., et al. 2010, ApJ, 713, 1073

Schaefer, B. E. \& Pagnotta, A. 2012, Natur, 431, 164

Schmidt, G. D., Hoard, D. W., Szkody, P., Melia, F., Honeycutt, R. K. \& Wagner, R. M. 1999, ApJ, 525, 407

Schmidt, G. D., Harris, H. C., Liebert, J., et al. 2003, ApJ, 595. 1101

Shen, K. J. \& Bildsten, L. 2007, ApJ, 660, 1444

Silverman, J. M., Ganeshalingam, M., Li, W., et al. 2011, MNRAS, 410, 585

Soker, N. 2011,  arXiv1109.4652S

Sokoloski, J. L. \& Beldstin, L. 1999, ApJ, 517, 919

Spitzer, L. 1962, Physics of Fully Ionized Gases, (New York: John Wiley and Sons)

Verbunt, F., \& Zwaan, C. 1981, A\&A, 100, L7

Wang, B., Li, X.-D., \& Han, Z. 2010, MNRAS, 401, 2729

Webbink, R. F., 1979, in van Horn H.-M., Weidemann V., eds, IAU Colloq.
53: White Dwarfs and Variable Degenerate Stars. Rochester University,
p. 426

Webbink, R. F. 1984, ApJ, 277, 355

Wheeler, J. C. 2012, ApJ, 758, 123

Woosley, S. E. \& Wallace, R. K. 1982, ApJ, 258, 716

Yaron, O., Prialnik, D., Shara, M. M. \& Kovetz, A. 2005, ApJ, 623, 398

Yoon, S.-C., \& Langer, N. 2004, A\&A, 419, 623

Yungelson, L. R., Livio, M., Truran, J. W., Tutukov, A. \& Fedorov, A. 1996, ApJ,
446, 890


\begin{figure}
\centering
\includegraphics[totalheight=3.5in,width=4.5in]{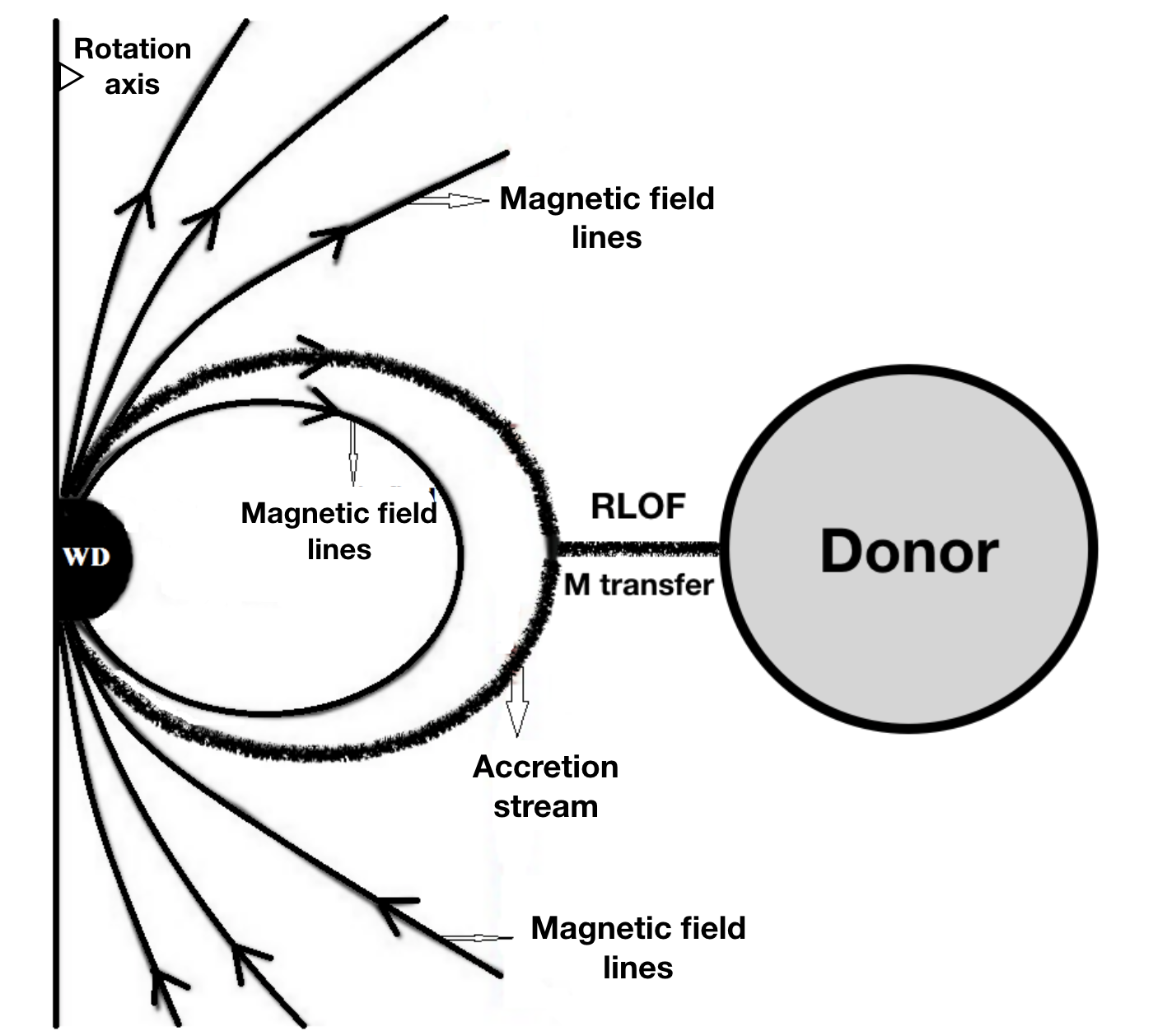}
\caption{The stream-like accretion along the magnetic lines of the WD in the highly magnetized WD binary.}
\label{fig:1}
\end{figure}

\clearpage



\begin{figure}
\centering
\includegraphics[totalheight=5.8in,width=6.5in]{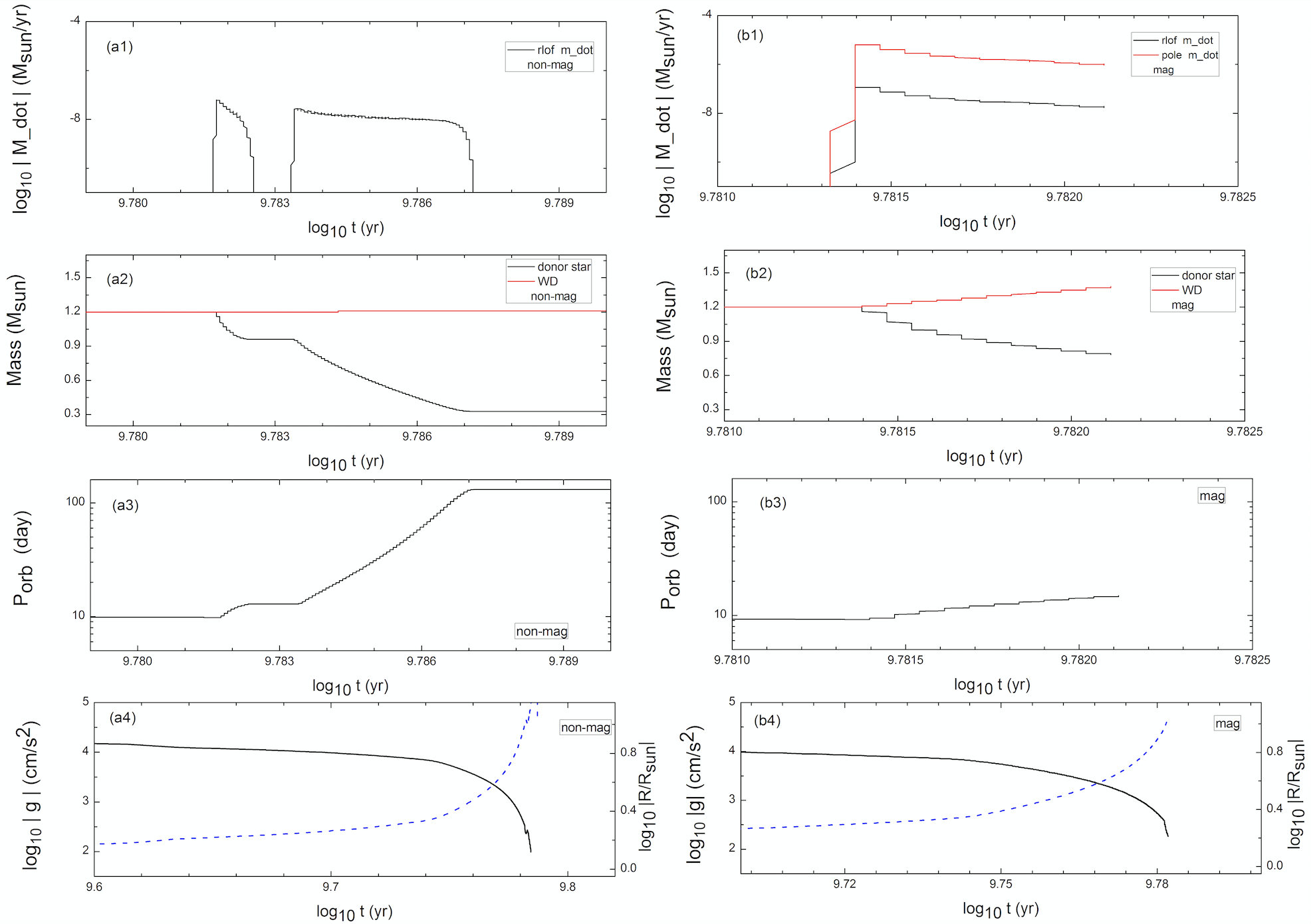}
\caption{Detailed evolution of WD binaries (evolutions of the WD/donor mass, orbital period, mass transfer, surface gravity and radius with time) by the MESA code. The two binaries have same initial conditions except the case of non-magnetic (left panels) and the magnetic WD (B$= 1.1\times10^8$ G: right panels). The initial masses of the WDs and donor stars are 1.2 $M_\sun$ with the initial orbital periods of 10 days. The blue dashed lines are the radius evolutions of the donors with time.}
\label{fig:1}
\end{figure}

\clearpage



\begin{figure}
\centering
\includegraphics[totalheight=5.8in,width=6.5in]{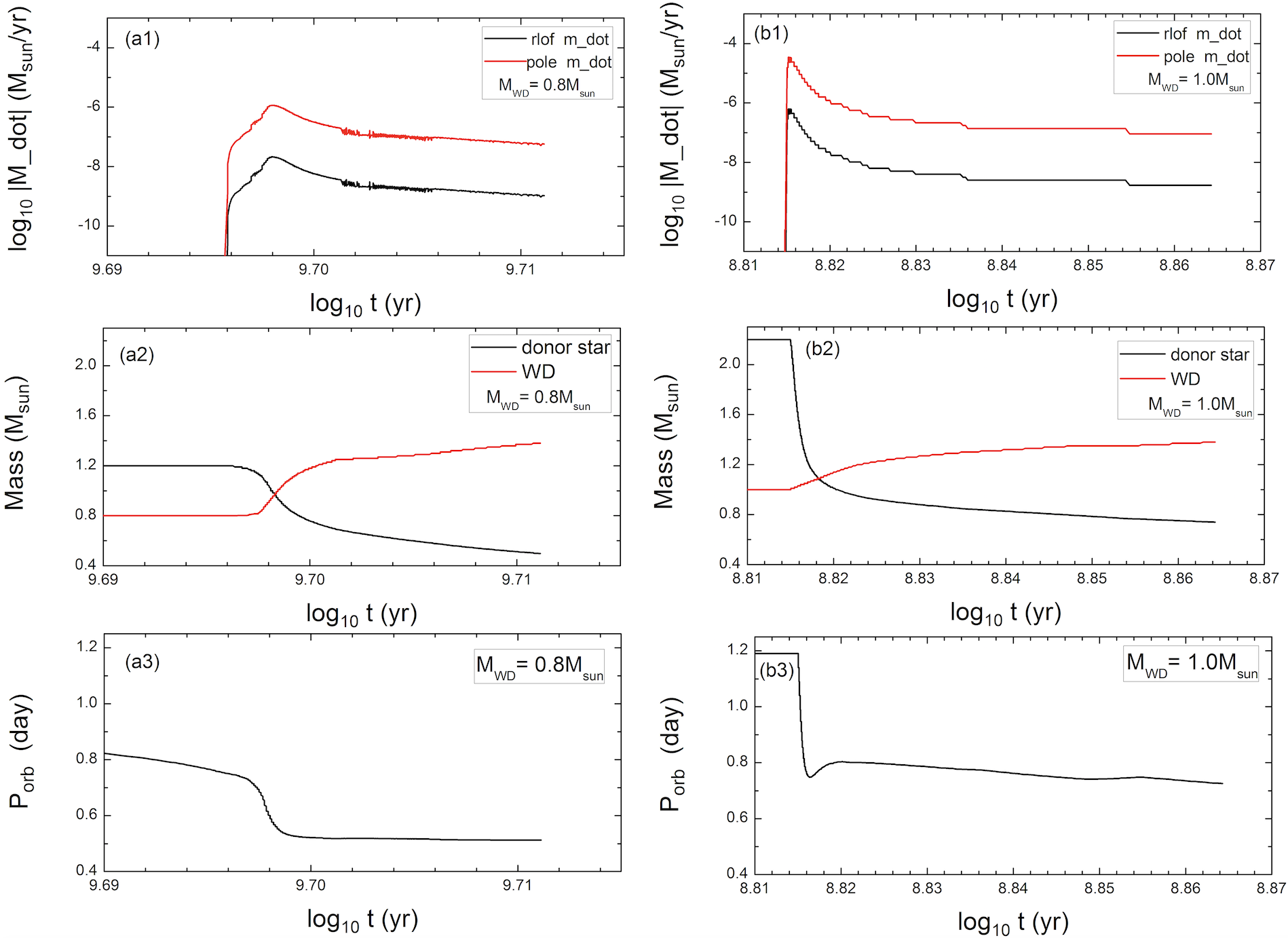}
\caption{Detailed evolution of WD binaries (evolutions of the WD/donor mass, orbital period and mass transfer with time)  by the MESA code.  The left panels for the binary: the WD initial mass is 0.8 $M_\sun$ with B$= 1.53\times10^8$ G, and the initial mass of the donor is 1.2 $M_\sun$ with the initial orbital period of 1 day. The right panels for the binary: the WD initial mass is 1.0 $M_\sun$ with B$= 1.1\times10^8$ G, and the initial mass of the donor and initial orbital period are 2.2 $M_\sun$ and 1.2 days, respectively.}
\label{fig:1}
\end{figure}

\clearpage

\begin{figure}
\centering
\includegraphics[totalheight=2.9in,width=3.0in]{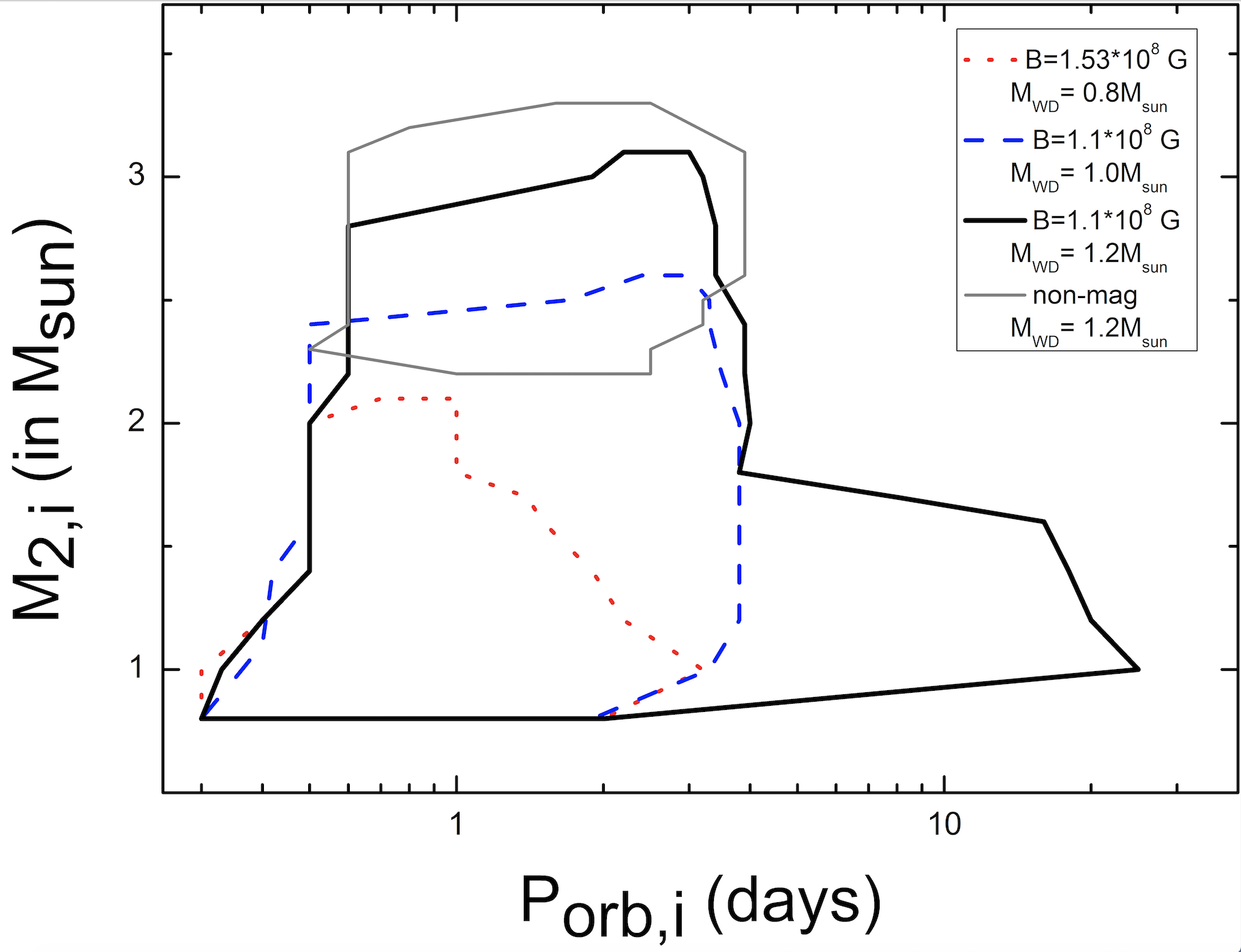}
\includegraphics[totalheight=2.9in,width=3.0in]{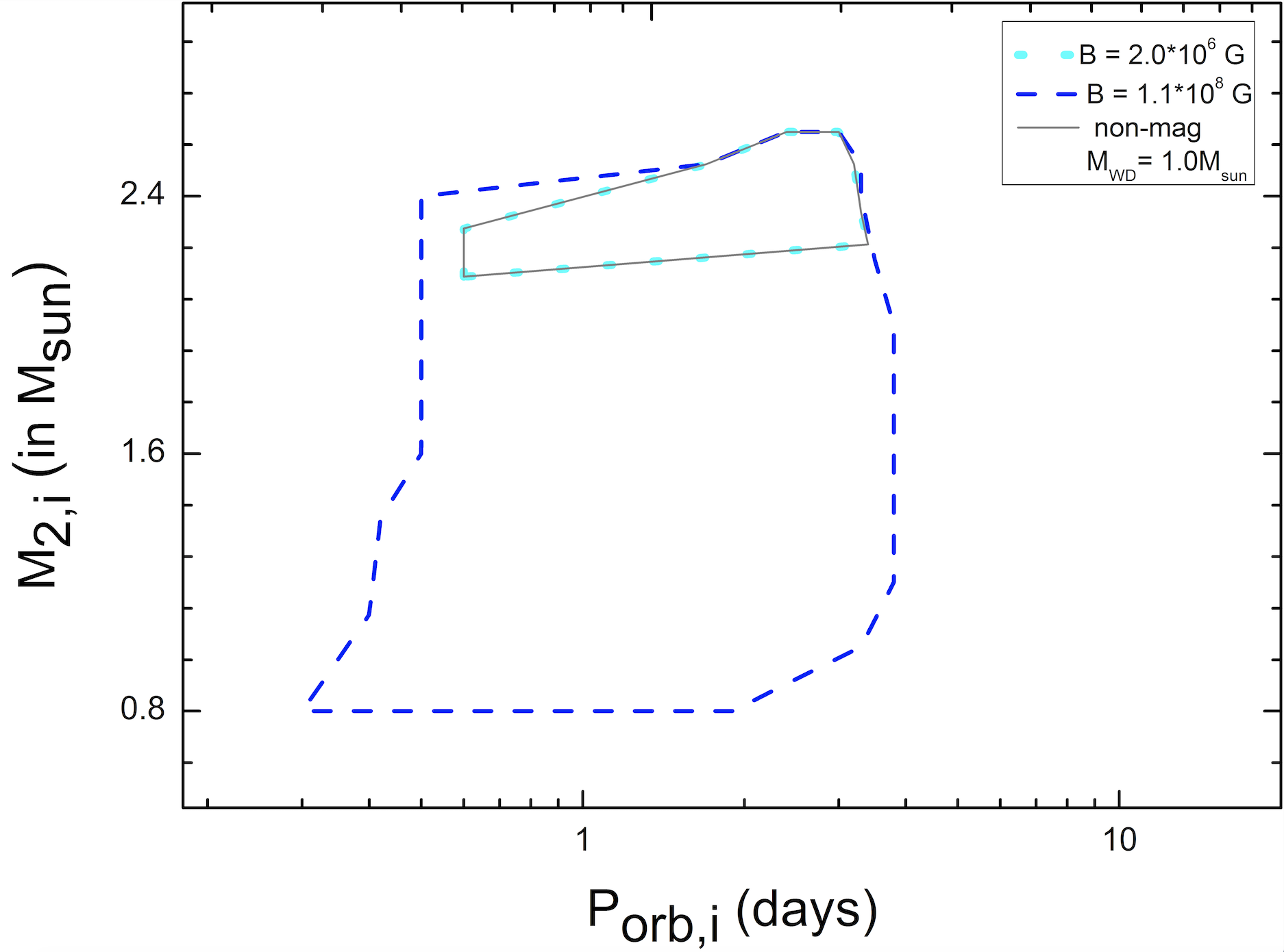}
\caption{Initial parameter space distributions for SNe Ia. The left panel: The red dotted line, blue dashed line and black solid line are for the binaries with $\rm M_{\rm WD, i} = 0.8\rm M_\sun$ with B $=1.53\times10^8$ G,  $\rm M_{\rm WD, i} = 1.0\rm M_\sun$ with B $=1.1\times10^8$ G, and $\rm M_{\rm WD, i} = 1.2 \rm M_\sun$ with B $=1.1\times10^8$ G, respectively. The gray solid line is for the case of non-magnetic WD with  $\rm M_{\rm WD, i} = 1.2 \rm M_\sun$. The right panel: The blue dashed, cyan-sick-dotted line and gray-thin-solid line for the binaries $M_{\rm WD, i} = 1.0\rm M_\sun$ under B$= 1.1\times10^8$ G, B$= 2.0\times10^6$ G and non-magnetic field cases, respectively.}
\label{fig:1}
\end{figure}

\clearpage


\begin{figure}
\centering
\includegraphics[totalheight=2.7in,width=3.4in]{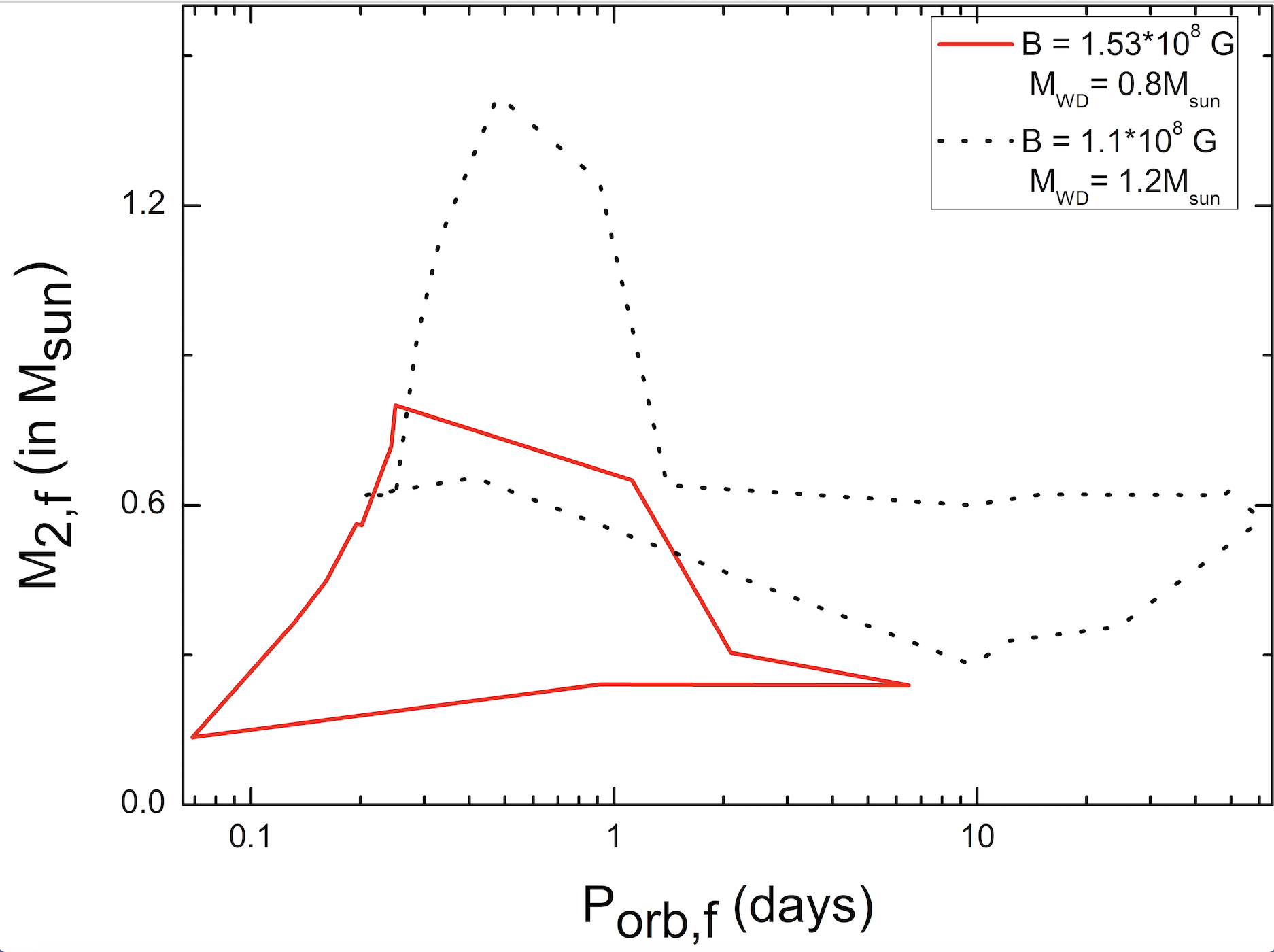}
\includegraphics[totalheight=2.7in,width=3.4in]{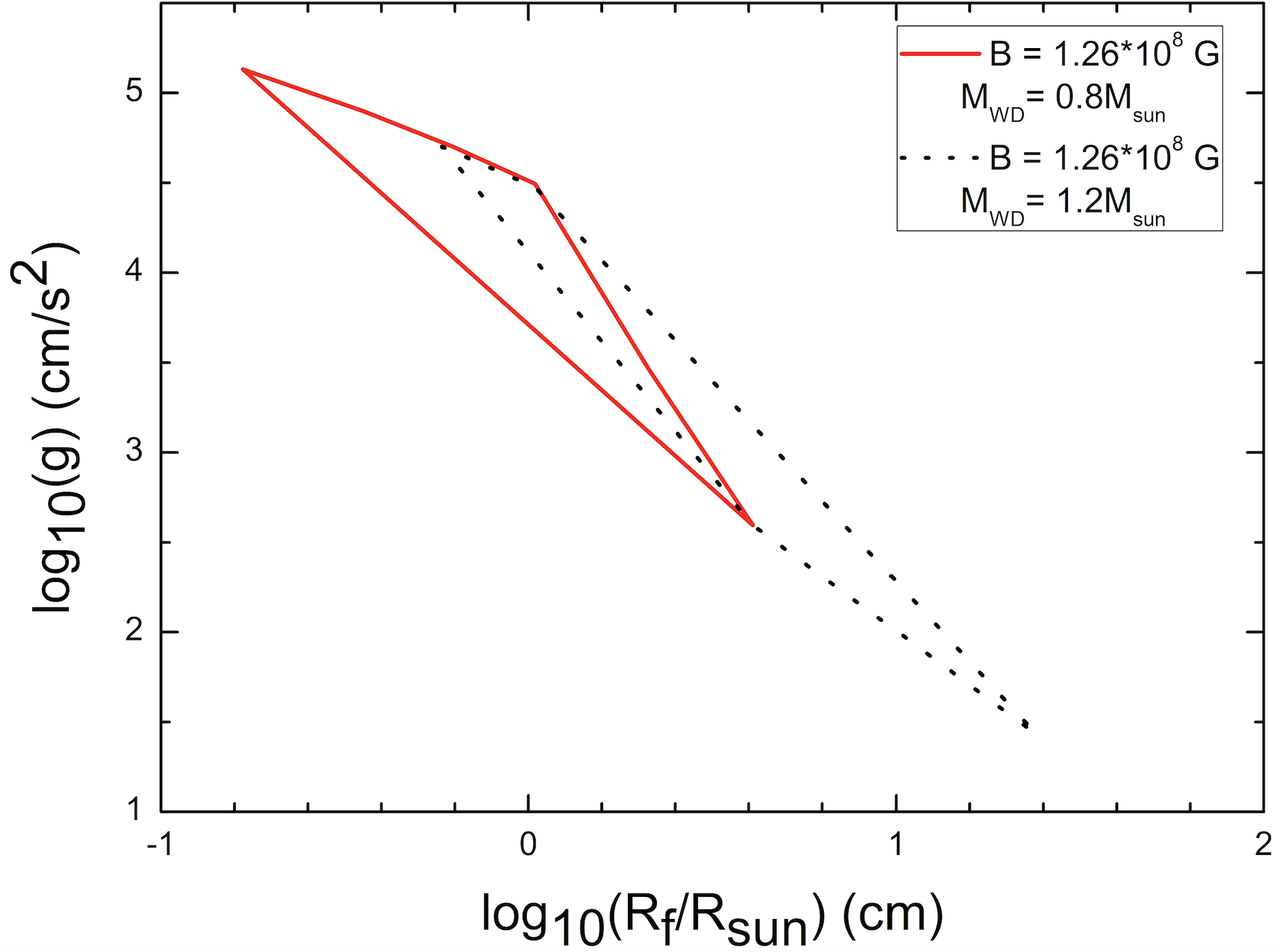}
\includegraphics[totalheight=2.7in,width=3.5in]{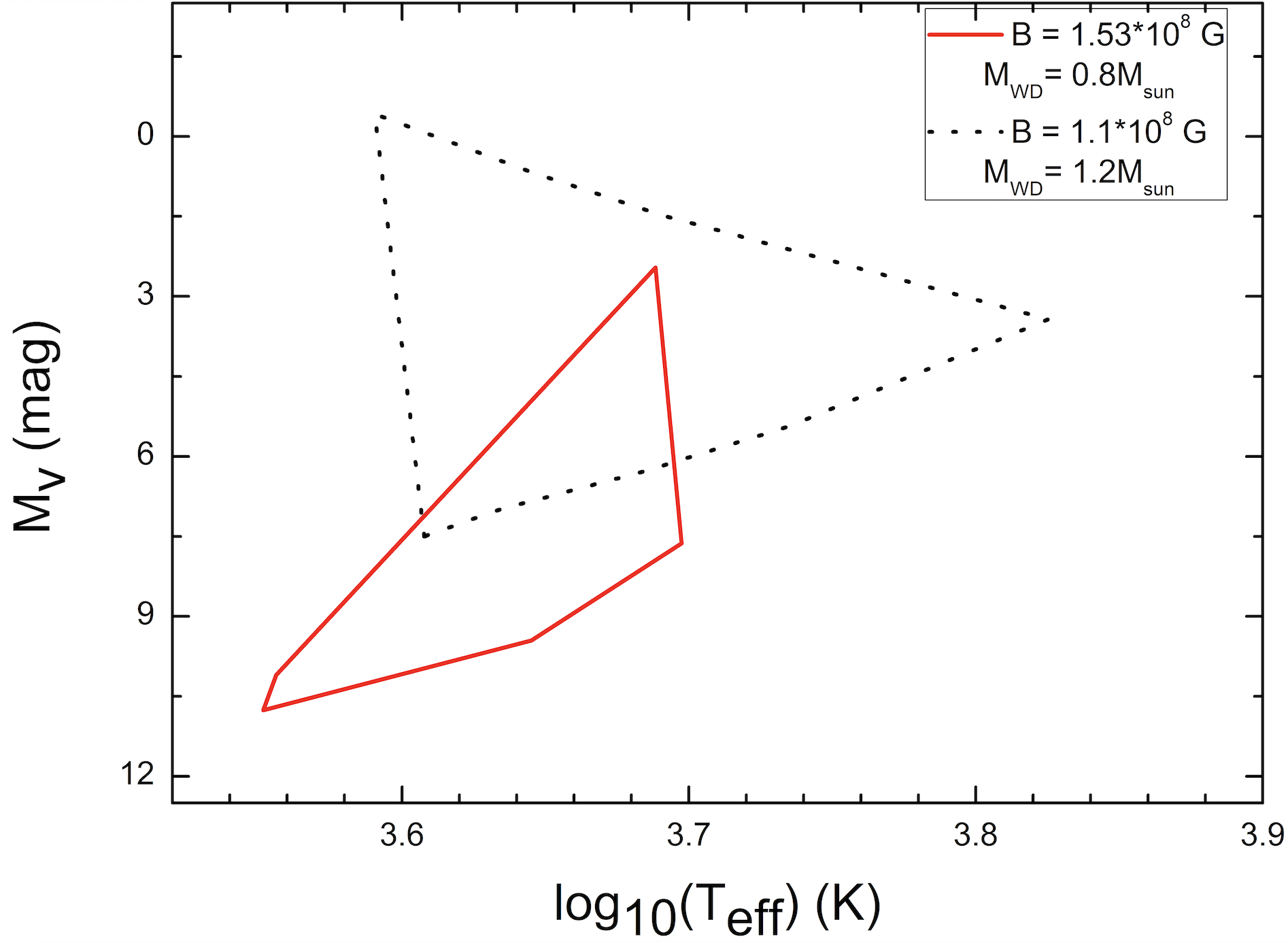}
\caption{The distributions of the final mass of the donor stars -- orbital period of the WD binaries, radius -- surface gravity of the donor stars and effective temperature -- absolute magnitude of the donor stars when the WDs ($\rm M_{\rm WD, i} = 0.8$ and $1.2 \rm M_\sun$) evolve to the Chandrasekhar limit mass. The black dotted and red solid lines are results from the binaries with $M_{\rm WD, i} =1.2,\, \&\, 0.8\,\rm M_\sun$, respectively.}
\label{fig:1}
\end{figure}



\clearpage

\end{document}